\documentclass[12pt]{article}

\catcode`\@=11
\global\arraycolsep=2pt
\usepackage{amsbsy,amssymb,latexsym,amsfonts,amsmath}
\usepackage{graphicx,color}

\begin{document}
\begin{flushright}
\parbox{4.2cm}
{RUP-18-33}
\end{flushright}

\vspace*{0.7cm}

\begin{center}
{ \Large Very Special $T\bar{J}$ deformed CFT}
\vspace*{1.5cm}\\
{Yu Nakayama}
\end{center}
\vspace*{1.0cm}
\begin{center}

Department of Physics, Rikkyo University, Toshima, Tokyo 171-8501, Japan

\vspace{3.8cm}
\end{center}

\begin{abstract}
We study a very special class of $T\bar{J}$ deformations of conformal field theories in two dimensions. While the deformations break the Lorentz symmetry, they preserve the twisted Lorentz symmetry. The resulting theory has right-moving Virasoro as well as left-moving translation and left-moving (chiral) scale symmetry without left-moving special conformal symmetry (nor left-moving Virasoro symmetry). As in the original $T\bar{J}$ deformations, they may be regarded as an operator dependent non-local change of coordinates. 
We show concrete examples based on worldsheet string theory and discuss how the non-unitary nature enables us to circumvent various no-go theorems. 
 %They are left conformal and right scale invariant, but the lack of right-moving Virasoro symmetry makes them in appropriate for the consistent string background.
\end{abstract}

\thispagestyle{empty} 

\setcounter{page}{0}

\newpage


%-----------------------------------------

%\pacs{}
%\keywords{Suggested keywords}%Use shokeys class option if keyword
%\maketitle

%%%%%%%%%%%%%%%%%%%%%%%%%%%%%%%%%%%%%%%%%%%%%%%%%%%%
\section{Introduction}
Understanding ultraviolet completions of a quantum field theory is an important issue. In particular, if the theory under consideration is power-counting non-renormalizable, the question is non-perturbative in nature and it is quite challenging. The better understanding of the ultraviolet completion would lead to deeper understanding of asymptotic safety scenario or non-perturbative ultraviolet fixed points proposed in the literature.
Recently, there has been some progress in this direction in two-dimensional conformal field theories perturbed by integrable non-renormalizable interactions. 

The first example is the so-called $T\bar{T}$ deformations \cite{Zamolodchikov:2004ce} of conformal field theories. In \cite{Smirnov:2016lqw}\cite{Cavaglia:2016oda}, they proposed that the deformations are integrable with a particular renormalization prescription, deriving the universal formula for the spectrum on the torus. It shows the Hagedorn-like ultraviolet behavior and the ultraviolet completion would be non-local, but some important features of the local quantum field theories still remain. More recently there has been a study in relation to the holography as well as in (little) string theory \cite{Giveon:2017nie}\cite{Giveon:2017myj}\cite{Asrat:2017tzd}\cite{Giribet:2017imm}\cite{Kraus:2018xrn}\cite{Cottrell:2018skz}\cite{Baggio:2018gct}\cite{Babaro:2018cmq}\cite{Babaro:2018cmq}. Studies of the uniqueness of the deformed partition function that can be realized in random geometries are further discussed in \cite{Cardy:2018sdv}\cite{Dubovsky:2018bmo}\cite{Datta:2018thy}\cite{Aharony:2018bad}\cite{Bonelli:2018kik}\cite{Conti:2018tca}. The supersymmetric generalizations can be found in \cite{Baggio:2018rpv}\cite{SUSY}.

Another class of irrelevant deformations studied in the literature is the so-called $T\bar{J}$ deformations \cite{Guica:2017lia}. They break the Lorentz symmetry, but preserve the chiral or right-moving Virasoro symmetry. The ultraviolet completion with the Lorentz breaking may be of interest, for such theories can be regarded as a toy model of Lorentz breaking renormalizable completion of non-renormalizable Lorentz invariant field theories. There is another interest from the holography. The $T\bar{J}$ deformations are candidates for holographic duals with Warped AdS space or near horizon limit of extremal Kerr black holes \cite{Bzowski:2018pcy}\cite{Chakraborty:2018vja}\cite{Apolo:2018qpq}\cite{Aharony:2018ics}.

The way in which the $T\bar{J}$ deformations break the Lorentz symmetry is special because it has a certain remnant of null directions. This reminds us of the very special relativity studied in four dimensions. In their pursuit in particle phenomenology, Cohen and Glashow asked what would be subgroups of the Poincar\'e group that preserve a particular null direction, and they classified the resulting Lorentz symmetry breaking pattern \cite{Cohen:2006ky}\cite{Cohen:2006ir}. In particular, they argued that some subgroups (i.e. $HOM(2)$ or $SIM(2)$) of the Lorentz symmetry are  consistent with the constancy of the speed of light but cannot be realized in local quantum field theories. In \cite{Nakayama:2017eof}\cite{Nakayama:2018qcs}, we have generalized the classification by adding the special conformal symmetries. Furthermore, in \cite{Nakayama:2018fib}, we have found a way to circumvent the no-go argument by Cohen and Glashow to construct the local field realizations by using the twisted Lorentz symmetry to present models of $HOM(2)$ or $SIM(2)$ invariant very special conformal field theories.

Now we may want to ask the similar question in two dimensions. Are there any quantum field theories with the symmetry analogous to the very special conformal symmetry? In the language of Virasoro algebra, the question is if  there is any quantum field theory that preserves $\bar{L}_{0}$, $\bar{L}_{\pm 1}$, ${L}_0$, ${L}_1$ without ${L}_{-1}$, which is similar to the $HOM(2)$ or $SIM(2)$ very special conformal symmetry in four dimensions. We, however, have a couple of no-go theorems here. In a sense, in two dimensions, (chiral) scale invariance implies conformal invariance, so it is quite non-trivial to find scale invariant field theories without conformal invariance. Our idea is motivated by the above-mentioned construction in four dimensions. We use the twisted Lorentz symmetry to obtain such theories. By construction, the theory becomes non-unitary to implement the twisted Lorentz algebra, but we show that such theories can arise by very special $T\bar{J}$ deformations of worldsheet string theory. This reveals the fact that we need more than finiteness or scale invariance to assure the consistency of the worldsheet string theory.

In this paper, we often use the light-cone notations in two-dimensional field theories. Our convention is
\begin{align}
x^\pm = \frac{1}{\sqrt{2}}(t\pm x)
\end{align}
and we call that the fields only dependent on $x^-$ as left-moving and fields only dependent on $x^+$ as right-moving. For example, a component of the (traceless) energy-momentum tensor  $T^{+}_{\ -} = -T_{- -}$ satisfying $\partial_+ T^{+}_{\ -} = 0$ is left-moving. In the Euclidean setup $T_{--} = -T^{+}_{\ -}$ is identified with the holomorphic energy-momentum tensor $T(z)$.

\section{Emergence of Conformal Symmetry}
The first question we would like to address is what is the subgroup of (global) conformal symmetry that can be realized in local quantum field theories. Assuming the space-time translation (i.e. $L_1$ and $\bar{L}_1$), we may add the Lorentz symmetry $J = L_0 - \bar{L}_0$, scale symmetry $D = L_0 + \bar{L}_0$, a linear combination thereof or both. A particularly interesting choice is to consider the chiral scale transformation $\bar{L}_0$ without ${L}_0$. This is known as the chiral scale invariance. Adding the chiral special conformal transformation further, we are led to the very special conformal symmetry. The case with $\bar{L}_{0}$, $\bar{L}_1$, $\bar{L}_{-1}$,  ${L}_{1}$ without ${L}_0$ and ${L}_{-1}$ has been studied in the literature \cite{Hofman:2011zj}. 
Our focus in this paper is the case with $\bar{L}_{0}$, $\bar{L}_{\pm 1}$, ${L}_0$, ${L}_1$ without ${L}_{-1}$. Are there any such theories? We actually have some no-go theorems, which we now review.

In two space-time dimensions, it is typical that the conformal symmetry emerges in quantum field theories with smaller space-time symmetries. In order to better understand the very special $T\bar{J}$ deformations of conformal field theories that do not show such enhancement, which are examples of very special conformal field theories, we first review the theoretical origin of the symmetry enhancement. We emphasize that the causality and unitarity plays a significant role in the emergence of the conformal symmetry.

Let us begin with the claim that in Poincar\'e invariant field theories scale invariance implies conformal invariance in two dimensions \cite{Polchinski:1987dy}. We assume that the space-time symmetry is realized by local conserved current i.e. energy-momentum tensor $T_{\mu\nu}$ in the canonical manner. 
The Poincar\'e invariance together with the scale invariance demands the conservation of the energy-momentum tensor and the special properties of its trace:
\begin{align}
\partial_+ T^{+}_{\ -} + \partial_- T^{-}_{\ -}& = 0 \cr
\partial_+ T^{+}_{\ +} + \partial_- T^{-}_{\ +}& = 0 \cr
T^{-}_{\ -} + \partial_+ s^+ +  \partial_- s^- &= 0 \cr
T^{+}_{\ +} + \partial_+\tilde{s}^+ + \partial_- \tilde{s}_- &=0 \ .
\end{align}
Here the energy-momentum tensor can be improved 
\begin{align}
T^{-}_{\ -} &\to T^{-}_{\ -} + \partial_+ l^+ \cr
T^{+}_{\ -} &\to T^{+}_{\ -} - \partial_- l^+ \cr
s^+ &\to s^+ - l^+ 
\cr
T^{+}_{\ +} &\to T^{+}_{\ +} + \partial_- l^- \cr
T^{-}_{\ +} &\to T^{-}_{\ +} - \partial_+ l^- \cr
\tilde{s}^- &\to \tilde{s}^- - l^- 
\end{align}
with arbitrary $l^+$ and $l^-$ so that $s^+$ and $\tilde{s}^-$ are actually arbitrary. 
In particular we can choose $s^+ = \tilde{s}^+$ and $s^- = \tilde{s}^-$ so that the energy-momentum tensor is symmetric $T^{-}_{\ -} = T^{+}_{\ +}$ as is known as the Belinfante prescription. 

Assuming the energy-momentum tensor $T^{+}_{\ -}$ has canonical scaling dimensions i.e. $(\Delta,\bar{\Delta})= (2,0)$, we have
\begin{align}
\langle T^{+}_{\ -}(x) T^{+}_{\ -}(0) \rangle  = \frac{c}{(x^-)^4} 
\end{align}
which means 
\begin{align}
\langle \partial_+ T^{+}_{\ -}(x) \partial_+ T^{+}_{\ -}(0) \rangle  = 0 \label{ddd}
\end{align}
up to possible contact terms. This implies $\partial_+ T^{+}_{\ -}(x)|0\rangle =0$ acting on the Lorentz invariant vacuum $|0\rangle$ from the unitarity. Then ``the state operator correspondence" valid in relativistic quantum field theories implies $\partial_+T^{+}_{\ -}(x) = 0$ as an operator identity. With this chiral conservation, we can construct the left-moving Virasoro generators $L_{-n+1}$
from
\begin{align}
\partial_+ ( (x^{-})^n T^{+}_{\ -}(x)) = 0 \ 
\end{align}
including the special conformal transformation with $n=2$. The same argument also applies to the right-moving sector: 
\begin{align}
\partial_- ( (x^{+})^n T^{-}_{\ +}(x)) = 0 \ . 
\end{align}
The conservation of the energy-momentum tensor further implies 
\begin{align}
\langle T^{-}_{\ -} (x) T^{-}_{\ -} (0) \rangle = \langle T^{+}_{\ +} (x) T^{+}_{\ +} (0) \rangle  = 0 \ , \label{ccc}
\end{align}
which follows taking $\partial_-$ derivatives of the left and side of \eqref{ccc} and comparing it with \eqref{ddd}. Then the unitarity requires that the energy-momentum tensor is traceless. This completes the argument that scale invariance implies conformal invariance in Poincar\'e invariant field theories in two dimensions.

The assumption to derive the conformal symmetry above can be further relaxed. For example, in \cite{Sibiryakov:2014qba} they argued that the Lorentz symmetry can be replaced by a causality (e.g. the commutators of local operators vanish outside of the certain ``light-cone"). On the other hand, in \cite{Hofman:2011zj}, they replaced the scale symmetry $D$ and the Lorentz symmetry $J$ by the chiral scale symmetry $D-J$ alone (without $D+J$).

In the case of chiral scale invariance, the space-time translation symmetry and the chiral scale symmetry demands the conservation of the energy-momentum tensor as 
\begin{align}
\partial_+ T^{+}_{\ -} + \partial_- T^{-}_{ \ -} &= 0 \cr
\partial_+ T^{+}_{\ +} + \partial_- T^{-}_{\ +} & = 0 \cr
T^{+}_{\ +} + \partial_+ s^+ + \partial_- s^- & = 0 
\end{align}
with arbitrary $s^-$ due to the improvement: 
\begin{align}
T^{+}_{\ +} &\to T^{+}_{\ +} + \partial_- l^- \cr
T^{-}_{\ +} &\to T^{-}_{\ +} - \partial_+ l^- \cr
{s}^- &\to {s}^- - l^- 
\end{align}
with arbitrary $l^-$.

By setting $s^- =0$, and assuming the canonical (chiral) scaling dimension $\bar{\Delta} = 0$ of $s^+$, we have
\begin{align}
\langle s^+(x) s^+(0) \rangle = s(x^-) 
\end{align}
so
\begin{align}
\langle \partial_+ s^+(x) \partial_+ s^+ (0) \rangle = 0 \ .
\end{align}
If we assume unitarity and the state operator correspondence,\footnote{One way to assure the state operator correspondence is to assume the causality by demanding commutators of local operators vanish outside of a certain ``light-cone", and use the analogue of the Reeh-Schlieder theorem valid here.}
 we must have 
$\partial_+ s^+(x) = 0$, leading to $T^{+}_{\ +} = 0$. The conservation of the energy-momentum tensor then requires $\partial_- T^{-}_{ \ +}$ = 0. Therefore, we have the right-moving Virasoro generators from 
\begin{align}
\partial_- ( (x^{+})^n T^{-}_{\ +}(x)) = 0 \ 
\end{align}
including the right-moving special conformal transformation with $n=2$.

Considering the left-mover, the canonical (chiral) scaling dimension $\bar{\Delta} = 0$ of  $T^{+}_{\ -} (x)$ demands
\begin{align}
\langle T^{+}_{\ -} (x) T^{+}_{\ -}(0) \rangle = t(x^-) \ ,
\end{align}
which  implies 
\begin{align}
\langle \partial_+ T^{+}_{\ -} (x) \partial_+ T^{+}_{\ -}(0) \rangle = 0 \ .
\end{align}
If we assume the unitarity and the state operator correspondence,
 we must have 
$\partial_+ T^+_{\ -}(x) = 0$ as an operator identity. As long as $T^{+}_{\ -}$ is non-zero, we can now construct the enhanced left-moving Virasoro generators
\begin{align}
\partial_+ ( (x^{-})^n T^{+}_{\ -}(x)) = 0 \ .
\end{align}
If  $T^{+}_{\ -} =0$, then  $T^{-}_{\ -}$ must be non-zero and the conservation demands $\partial_- T^{-}_{\ -} = 0$.  Then we can construct the right-moving current algebra
\begin{align}
\partial_- ((x^{+})^n T^{-}_{\ -}(x)) = 0 
\end{align}
whose zero mode $(n=0)$ generates the left-moving translation.
This completes the argument for the enhancement of symmetries in chiral scale invariant field theories in two dimensions.

\section{General structure of very special $T\bar{J}$ deformations}
In this section, we propose a general recipe to construct very special conformal field theories with the right-moving Virasoro, left-moving translation and left-moving twisted scale symmetry from the very special $T\bar{J}$ deformations. To appreciate the very special nature, we begin with the ordinary $T\bar{J}$ deformations of conformal field theories.

Let us consider a two-dimensional conformal field theory with a conserved $U(1)$ current $J$. The energy-momentum tensor is taken to be traceless
\begin{align}
\partial_+ T^{+}_{\ -} (x^-) = 0 \cr
\partial_- T^{-}_{\ +} (x^+) = 0 
\end{align}
and the current is taken to be purely left-moving $(x^-)$ or right-moving $(x^+)$ for simplicity.
\begin{align}
\partial_+ J^+(x^-) = 0 \cr
\partial_- J^{-}(x^+) = 0 \ .
\end{align}
This is the case when the theory is unitary and has a discrete spectrum. We will assume $J^{-}$ does not vanish. Otherwise we could switch the left-mover and right-mover in the following.

Let us now deform the conformal field theory by introducing the irrelevant Lorentz breaking term in the action
\begin{align}
\delta S = \mu \int dx dt J^{-}(x^+) T^{+}_{\ -}(x^-) \ 
\end{align}
with an appropriate ultraviolet completion.
This is called the $T\bar{J}$ deformation in the literature. Note that the deformation breaks the Lorentz invariance in a manner such that it preserves a particular null direction $x^+$.

Treating the deformation as a perturbative series with respect to $\mu$, the energy-momentum tensor as well as the current conservation are modified 
\begin{align}
\partial_+ T^{+}_{\ -} + \partial_- (\mu J^- T^{+}_{\ -}) &= 0 \cr
\partial_- T^-_{\ +} &= 0 \cr
\partial_- J^-  &= 0 \cr
\partial_+ J^+ + \partial_-(\mu J^+ J^-) & = 0
\end{align}
at the first order in $\mu$. One may therefore observe that $\mu J^- T^{+}_{\ -}$ is identified with $T^{-}_{\ -}$ so that the left-moving translation is generated by the energy-momentum tensor with the modified conservation. At this order, on the other hand,  we see that the right-moving energy-momentum tensor is intact and the theory preserves the right-moving Virasoro symmetry from the conservation
\begin{align}
\partial_- ((x^+)^n T^-_{\ +}(x) ) = 0 \ . 
\end{align}
Note that the left-moving (chiral) scale and the left-moving special conformal transformation are (apparently) broken.

There is an alternative look at the $T\bar{J}$ deformations as an operator dependent non-local change of coordinates. The deformation, at the first order in $\mu$ is induced by the operator dependent left-moving coordinate transformation
\begin{align}
x^+ &\to x^+ \cr
x^- &\to x^- - \mu \int^{x^+} d\tilde{x}^{+} J^{-} (\tilde{x}^+) \   
\end{align}
to the original conformal field theories. In the simple situation in which the left-moving sector and right-moving sector effectively decouples, the coordinate transformation can be performed explicitly and the deformations become integrable. In relation, \cite{Bzowski:2018pcy} observed that the deformed theory may possess the infinitely many non-local charges
\begin{align}
\mathcal{L}_n  &= \int dx \left(x^- -\mu \int^{x^+} d\tilde{x}^{+} J^{-} (\tilde{x}^+) \right)^n (T^{+}_{\ -} - T^{-}_{\ -}) \cr 
 &= \int dx \left(x^- -\mu \int^{x^+} d\tilde{x}^{+} J^{-} (\tilde{x}^+) \right)^n (1-\mu J^{-}) T^{+}_{\ -}
\end{align}
which can be understood as the fate of the original right-moving Virasoro charges after the non-local operator dependent coordinate changes.

One may also regard the $T\bar{J}$ deformations as an operator dependent non-local gauge transformations for the symmetry generated by the $\bar{J}$ current. Under the gauge transformations with the parameter $\mu \int^{x^-} d\tilde{x}^-  T^{+}_{\ -}(\tilde{x}^-)$, the first oder deformations of the action is given by the $T\bar{J}$ deformations.
 However, the chiral nature of $T^{+}_{\ -}(\tilde{x}^-)$ is lost, unlike $J^{-} (\tilde{x}^+)$, at the first order in perturbation, so this way of understanding the  $T\bar{J}$ deformations may not lead to the integrable structure beyond the first order in $\mu$.

In this paper, we are mostly interested in local charges. 
Comparing the situation with the claim in  \cite{Hofman:2011zj}, there is a small puzzle here because they claim that the local symmetry must be enhanced. To see the puzzle more explicitly, let us study the perturbed two-point functions
\begin{align}
\langle T^+_{\ -}(x) T^{+}_{\ -} (y) \rangle &= \frac{c}{(x^- - y^-)^4} + \mu^2 t(x-y) +O(\mu^3) \cr
\langle T^{-}_{\ -}(x) T^{-}_{\ -}(y) \rangle &= \mu^2 s(x-y) + O(\mu^3) \cr 
\langle J^{-} ({x}) J^{-} ({y}) \rangle & = \frac{k_J}{(x^+-y^+)^2} + O(\mu) \ .  
\end{align}
As we saw, if the chiral scale invariance holds in a strict sense, $t(x-y)$ is a function of $x^- - y^-$ only and does not depend on $x^+-y^+$, so taking derivatives with respect to $x^+$ and $y^{+}$ and using the energy-momentum tensor conservation, we obtain
\begin{align}
\langle \partial_+ T^+_{\ -}(x) \partial_+ T^{+}_{\ -} (y) \rangle = 
\langle \partial_- T^{-}_{\ -}(x) \partial_- T^{-}_{-}(y) \rangle  = 0 \label{HSp}
\end{align}
possibly up to contact terms. 

On the other hand, the explicit computation gives
\begin{align} 
\langle T^{-}_{\ -}(x) T^{-}_{-}(y) \rangle &= \mu^2 \langle  J^- T^{+}_{\ -}(x)  J^- T^{+}_{\ -}(y) \rangle  \cr
& = \frac{\mu^2 k_J c}{(x^- -y^-)^4 (x^+ - y^+)^2}  + O(\mu^3) 
\end{align}
at the first non-trivial order in $\mu$. We see that \eqref{HSp} does not hold unless $c$ or $k_J$ vanishes. Integrating the conservation equation twice, we obtain
\begin{align}
t(x-y) \sim c k_J \frac{ \log (x^+-y^+)}{(x^--y^-)^6}
\end{align}
which does depend on $x^+ - y^+$ in a logarithmic manner, and the right-moving scale symmetry is anomalously broken. The problem arises from the bad ultra-violet behavior of the perturbative computations of 
\begin{align}
\langle T^+_{\ -}(x) T^{+}_{\ -} (y) \rangle \sim 
 \mu^2 \langle T^+_{\ -}(x) T^{+}_{\ -} (y) \int d^2z_1 J^-(z_1^+)T^{+}_{\ -}(z_1^-)  \int d^2z_2 J^-(z_2^+)T^{+}_{\ -}(z_2^-) \rangle_0 \ .
\end{align}
This is badly divergent, but one may take the derivatives $\partial_+$ to make it finite by using the prescription (with a proper contour deformation)
\begin{align}
\partial_+ \frac{1}{x^- -z^-} = \pi i \delta (x-y) \ . 
\end{align}
Indeed, this is how the modified conservation is obtained from the beginning. This however results in the anomalous violation of the chiral scale symmetry.

Let us now turn to a very special class of the $T\bar{J}$ deformations to preserve extra twisted Lorentz symmetry. The idea is to choose $\bar{J}$ as a null component of the right-moving $SL(2)$ current algebra so that in the original conformal field theory, we had the operator product expansions: 
\begin{align}
\bar{J}(x) \bar{J}(y) &= 0 \cr
\bar{J}(x) \bar{K}(y) & = \frac{\bar{J}(y)}{(x^+ - y^+)} \ .
\end{align}
Note that in order to have a null current, the theory must be non-unitary.

Under these extra assumptions, we deform the original conformal field theories by the twisted marginal deformations
\begin{align}
\delta S = \mu \int dx dt J^{-}(x^+) T^{+}_{\ -}(x^-) \ .
\end{align}
We call it twisted marginal because the deformations had $(\Delta,\bar{\Delta}) = (2,1)$, but we may assign the twisted Lorentz charge  $(\tilde{\Delta},\bar{\Delta}) = (1,1)$ by using the fact that $J^-$ is charged under $K^-$ and defining $\tilde{\Delta} = \Delta - K^-$.

Treating the deformation as a perturbative series with respect to $\mu$, the energy-momentum tensor as well as the current conservation is modified 
\begin{align}
\partial_+ T^{+}_{\ -} + \partial_- (\mu J^- T^{+}_{\ -}) &= 0 \cr
\partial_- T^-_{\ +} &= 0 \cr
\partial_- J^-  &= 0 \cr
\partial_+ J^+ + \partial_-(\mu J^+ J^-)  & = 0
\end{align}
as before. In addition, since ${J}^-$ is charged under ${K}^-$, ${K}^-$ is no longer conserved after the $T\bar{J}$ deformation
\begin{align}
\partial_- K^- = \mu J^{-} T^{+}_{\ -} \ .
\end{align}

Our very special $T\bar{J}$ deformations can be regarded as a non-local operator dependent change of coordinates as in the original $T\bar{J}$ deformations, so the deformations are integrable, but it has further interesting features.
The point here is that the existence of the $SL(2)$ current algebra of the original conformal field theories enable us to construct the left-moving scale current or twisted Lorentz current from
\begin{align}
\partial_+(x^- T^{+}_{\ -}) + \partial_- (x^- T^{-}_{\ -} - K^-) = 0 \ 
\end{align}
with $T^{-}_{\ -} = \mu J^{-} T^{+}_{\ -}$.
Generically, $K^-$ is not a derivative of local operators, so the theory is not invariant under the left-moving special conformal transformation. If $K^{-} = \partial_- O + \partial_+ \tilde{O} $ with a local operator $O$, $\tilde{O}$ for some reasons, however, then we may further preserve the left-moving special conformal transformation
\begin{align}
\partial_+((x^-)^2 T^{+}_{\ -} + 2\tilde{O}) + \partial_- ((x^-)^2 T^{-}_{\ -} - 2x^- K^{-} + 2 O) = 0 \ .
\end{align}
Even in this case, there will be no left-moving Virasoro symmetry that is realized in a local manner.

Due to the existence of the extra twisted Lorentz symmetry, the ultraviolet behavior of the perturbation is improved. Indeed, the conservation of the $SL(2)$ charge dictates that the two-point functions of the energy-momentum tensor is not modified
\begin{align}
\langle \partial_+ T^+_{\ -}(x) \partial_+ T^{+}_{\ -} (y) \rangle = 
\langle \partial_- T^{-}_{\ -}(x) \partial_- T^{-}_{-}(y) \rangle  = 0 
\end{align}
in accordance with the argument in \cite{Hofman:2011zj} which we have reviewed in the previous section.  Note, however, that because of the lack of the unitarity, one cannot conclude that $\partial_+ T^+_{\ -}(x) =0$. Indeed it is non-zero and the -moving special conformal transformation is absent in more general correlation functions.

\section{Examples: very special $T\bar{J}$ deformed worldsheet theory}
In this section, we construct the very special $T\bar{J}$ deformed conformal field theories from free fermions. Since we use the non-unitary time-like fermions, the most familiar situation is in the worldsheet string theory. There we naturally have a fermion with a time-like kinetic term and the ghost fermions.
 The current $\bar{J}$ is then taken to be a null target space Lorentz current (possibly with the ghost directions included).

We begin with a free conformal field theory with three Majorana fermions
\begin{align}
S_0^{c=1} = \int dxdt \frac{i}{2} \left(\frac{1}{2}(b\partial_+c + c\partial_+b) + \psi \partial_+ \psi + \frac{1}{2}(\tilde{b}\partial_- \tilde{c}  + \tilde{c}\partial_- \tilde{b}) + \tilde{\psi} \partial_- \tilde{\psi} \right) \ .
\end{align}
Out of three fermions, one of them has a time-like kinetic term and we combine them as a $bc$ ghost system. Here we are going to choose the energy-momentum tensor so that the $bc$ ghost system has the Virasoro central charge $c=1$.

We now consider the very special $T\bar{J}$ deformation by adding 
\begin{align}
S_{\mu}^{c=1} = \int dx dt \left( \mu \tilde{b} \tilde{\psi}\left( \frac{1}{2} b\partial_- c + \frac{1}{2} c \partial_- b + \psi \partial_- \psi \right)  \right) \ . \label{TJd}
\end{align} 
Varying the deformed action, we derive the equations of motion
\begin{align}
0 &= \frac{i}{2} \partial_+ b + \mu \left(\frac{1}{2} \partial_-(\tilde{b} \tilde{\psi} b) + \frac{1}{2} (\tilde{b} \tilde{\psi} \partial_- b) \right) \cr
0 &= \frac{i}{2} \partial_+ c + \mu \left(\frac{1}{2} \partial_- (\tilde{b} \tilde{\psi} c) + \frac{1}{2} \tilde{b} \tilde{\psi} \partial_- c \right) \cr
0 & = \frac{i}{2} 2\partial_+ \psi + \mu \left(\partial_- (\tilde{b} \tilde{\psi} \psi ) + \tilde{b} \tilde{\psi} \partial_- \psi \right) \cr
0 &= \frac{i}{2} \partial_- \tilde{b} \cr
0 &= \frac{i}{2} \partial_- \tilde{c} + \mu \tilde{\psi} \left(\frac{1}{2} b \partial_- c + \frac{1}{2} c \partial_- b + \psi \partial_- \psi \right) \cr
0&= \frac{i}{2} (2\partial_- \tilde{\psi}) - \mu \tilde{b} \left(\frac{1}{2} b \partial_- c + \frac{1}{2} c \partial_- b + \psi \partial_- \psi\right) \ ,
\end{align}
some of which can be simplified as
\begin{align}
0 &= \frac{i}{2} \partial_+ b + \mu \tilde{b} \tilde{\psi} \partial_- b  \cr
0 &= \frac{i}{2} \partial_+ c + \mu \tilde{b} \tilde{\psi} \partial_- c  \cr
0 & = \frac{i}{2} \partial_+ \psi + \mu \tilde{b}\tilde{\psi} \partial_-\psi \cr
0 &= \frac{i}{2} \partial_- \tilde{b} \cr
0 &= \frac{i}{2} \partial_- \tilde{c} + \mu \tilde{\psi} \left(\frac{1}{2} b
 \partial_- c + \frac{1}{2} c \partial_- b + \psi \partial_- \psi \right) \cr
0&= \frac{i}{2} (2\partial_- \tilde{\psi}) - \mu \tilde{b} \left(\frac{1}{2} b \partial_- c + \frac{1}{2} c \partial_- b + \psi \partial_- \psi\right) \ .
\end{align}

Due to the translational invariance, we can always construct the canonical Noether energy-momentum tensor 
\begin{align}
T^{+}_{\ -} &= \frac{ \partial L} {\partial \partial_+ \Psi_i} \partial_- \Psi_i \cr
T^{-}_{\ -} &= \frac{\partial L} {\partial \partial_- \Psi_i} \partial_- \Psi_i - L \cr 
T^{+}_{\ +} & =  \frac{\partial L} {\partial \partial_+ \Psi_i} \partial_+ \Psi_i- L  \cr
T^{-}_{\ +} & =  \frac{\partial L} {\partial \partial_- \Psi_i} \partial_+ \Psi_i  \ ,
\end{align}
where the right derivative is assumed. They are conserved
\begin{align}
\partial_- T^{-}_{ \ -} + \partial_+ T^{+}_{\ -} &= 0 \cr
\partial_- T^{-}_{ \ +} + \partial_+ T^{+}_{\ +} &= 0 \ 
\end{align}
by using the equations of motion.

As we have discussed in the previous section, the currents are still conserved
\begin{align}
J^+ =  ib\psi   \cr
J^- = i \tilde{b} \tilde{\psi}  
\end{align}
with
\begin{align}
\partial_+ J^+ + \partial_- (2\mu \tilde{b}\tilde{\psi}b\psi) =  0 \cr
\partial_- J^- = 0 \ . 
\end{align}
Note that $J^-$ is still purely right-moving.
On the other hand, ${K}^- = i \tilde{b} \tilde{c}$ is broken by the interaction
\begin{align}
\partial_- {K}^- = -2 \mu \tilde{b} \tilde{\psi} \left(\frac{1}{2} (b \partial_- c + c \partial_- b) + \psi \partial_- \psi \right) \ .
\end{align}

Now the canonical energy-momentum tensor reads
\begin{align}
T^{-}_{\ +} &= \frac{i}{2} \left(\frac{1}{2}( \tilde{b} \partial_+ \tilde{c} + \tilde{c} \partial_+ \tilde{b}) + \tilde{\psi} \partial_+ \tilde{\psi} \right) +\mu \tilde{b} \tilde{\psi} \left(\frac{1}{2} (b \partial_+ c + c \partial_+ b) + \psi \partial_+ \psi \right)  \cr
T^{+}_{\ -} &= \frac{i}{2} \left(\frac{1}{2}( b \partial_- c  + {c} \partial_- {b}) +{\psi} \partial_-\psi  \right) \cr
T^{-}_{\ -} &= -\frac{i}{2} \left(\frac{1}{2}( b \partial_+ c  + {c} \partial_+ {b}) +{\psi} \partial_+ \psi  \right) \cr
T^{+}_{\ +} &= -\frac{i}{2} \left(\frac{1}{2}( \tilde{b} \partial_- \tilde{c} + \tilde{c} \partial_- \tilde{b}) + \tilde{\psi} \partial_- \tilde{\psi} \right) - \mu \tilde{b} \tilde{\psi} \left(\frac{1}{2} (b \partial_- c + c \partial_- b) + \psi \partial_- \psi \right) \ .
\end{align}
By using the equations of motion, we can simplify them as
\begin{align}
T^{-}_{\ +} &= \frac{i}{2} \left(\frac{1}{2}( \tilde{b} \partial_+ \tilde{c} + \tilde{c} \partial_+ \tilde{b}) + \tilde{\psi} \partial_+ \tilde{\psi} \right) \cr
T^{+}_{\ -} &= \frac{i}{2} \left(\frac{1}{2}( b \partial_- c  + {c} \partial_- {b}) +{\psi} \partial_-\psi  \right) \cr
T^{-}_{\ -} &= \mu \tilde{b} \tilde{\psi} \left(\frac{1}{2}(b\partial_- c + c\partial_-b) + \psi\partial_- \psi \right) 
\cr
T^{+}_{\ +} &= 0 \ .
\end{align}
Due to the chiral conservation $\partial_- T^{-}_{\ +} = 0$, the theory has the right-moving Virasoro symmetry from $\partial_- ((x^+)^n T^{-}_{\ +}) = 0$. In addition, the theory possesses the hidden (or twisted) left-moving invariance. To see this, we note that upon the use of the equations of motion
\begin{align}
T^{-}_{\ -} = -\frac{1}{2} \partial_- K^- \ , 
\end{align}
 we have the additional conservation 
\begin{align}
 \partial_+ (x^- T^{+}_{\ -} ) + \partial_- (x^- T^{-}_{\ -} + \frac{1}{2} K^{-})    =  0 \ ,
\end{align}
which generates the twisted left-moving scale transformation. Since ${K}^{-}$ is not (apparently) a derivative of something further,\footnote{One may try to bosonize the fermions, but the signature of the fermionic kinetic term makes it non-trivial.} 
we do not have the left-moving special conformal current (nor left-moving Virasoro current).

Indeed the action \eqref{TJd} is invariant under the two chiral rotations with the charges
\begin{align}
Q_L(\partial_-, \partial_+,b,c,\psi,\tilde{b},\tilde{c},\tilde{\psi}) &= (1,0,1/2,1/2,1/2,-1,1,0) \cr
Q_R (\partial_-, \partial_+,b,c,\psi,\tilde{b},\tilde{c},\tilde{\psi}) &= (0,1,0,0,0,1/2,1/2,1/2) \ , \label{twistedLorentz}
\end{align}
which is regarded as the twisted Lorentz transformation. Note that $Q_L$ is unbounded below unlike $Q_R$ due to the lack of unitarity.

Now let us consider the same action with a different energy-momentum tensor (i.e. $c=-26$)
\begin{align}
S_0^{c=-26} = \int dxdt \frac{i}{2} \left(\frac{1}{2}(2 b \partial_+ c - c\partial_+b  + \psi \partial_+ \psi + 2 \tilde{b} \partial_- \tilde{c} - \tilde{c} \partial_+ \tilde{b} + \tilde{\psi} \partial_- \tilde{\psi} \right) \ .
\end{align}
The action is the same up to the surface term, but the way we order the fields and derivatives change the canonical energy-momentum tensor by the improvement terms.  
Since we use the different energy-momentum tensor, the $T\bar{J}$ deformation is changed accordingly:
\begin{align}
S_\mu^{c=-26} = \int dx dt \left( \mu \tilde{b} \tilde{\psi}\left(2b \partial_- c - c \partial_-b  + \psi \partial_- \psi \right)  \right) \ .
\end{align}

Varying the action, we obtain the equations of motion
\begin{align}
0 &= \frac{i}{2} \partial_+ b + \mu \left(2\partial_-(\tilde{b} \tilde{\psi} b) - (\tilde{b} \tilde{\psi} \partial_- b) \right) \cr
0 &= \frac{i}{2} \partial_+ c + \mu \left(- \partial_- (\tilde{b} \tilde{\psi} c) +2 \tilde{b} \tilde{\psi} \partial_- c \right) \cr
0 & = \frac{i}{2} 2\partial_+ \psi + \mu \left(\partial_- (\tilde{b} \tilde{\psi} \psi ) + \tilde{b} \tilde{\psi} \partial_- \psi \right) \cr
0 &= \frac{i}{2} \partial_- \tilde{b} \cr
0 &= \frac{i}{2} \partial_- \tilde{c} + \mu \tilde{\psi} \left(2 b \partial_- c  - c \partial_- b + \psi \partial_- \psi \right) \cr
0&= \frac{i}{2} (2\partial_- \tilde{\psi}) - \mu \tilde{b} \left(2 b \partial_- c - c \partial_- b + \psi \partial_- \psi\right) \ ,
\end{align}
some of which can be simplified as
\begin{align}
0 &= \frac{i}{2} \partial_+ b + \mu \tilde{b} \tilde{\psi} \partial_- b  \cr
0 &= \frac{i}{2} \partial_+ c + \mu \tilde{b} \tilde{\psi} \partial_- c  \cr
0 & = \frac{i}{2} \partial_+ \psi + \mu \tilde{b}\tilde{\psi} \partial_-\psi \cr
0 &= \frac{i}{2} \partial_- \tilde{b} \cr
0 &= \frac{i}{2} \partial_- \tilde{c} + \mu \tilde{\psi} \left(2 b
 \partial_- c - c \partial_- b + \psi \partial_- \psi \right) \cr
0&= \frac{i}{2} (2\partial_- \tilde{\psi}) - \mu \tilde{b} \left(2 b \partial_- c - c \partial_- b + \psi \partial_- \psi\right) \ . 
\end{align}

The canonical energy-momentum tensor can be computed as 
\begin{align}
T^{-}_{\ +} &= \frac{i}{2} \left(2 \tilde{b} \partial_+ \tilde{c} - \tilde{c} \partial_+ \tilde{b} + \tilde{\psi} \partial_+ \tilde{\psi} \right) +\mu \tilde{b} \tilde{\psi} \left(2 b \partial_+ c - c \partial_+ b + \psi \partial_+ \psi \right)  \cr
T^{+}_{\ -} &= \frac{i}{2} \left(2b \partial_- c  - {c} \partial_- {b} +{\psi} \partial_-\psi  \right) \cr
T^{-}_{\ -} &= -\frac{i}{2} \left(2 b \partial_+ c  - {c} \partial_+ {b} +{\psi} \partial_+ \psi  \right) \cr
T^{+}_{\ +} &= -\frac{i}{2} \left(2 \tilde{b} \partial_- \tilde{c} - \tilde{c} \partial_- \tilde{b} + \tilde{\psi} \partial_- \tilde{\psi} \right) - \mu \tilde{b} \tilde{\psi} \left(2b \partial_- c - c \partial_- b + \psi \partial_- \psi \right) \ .
\end{align}
When $\mu=0$, the energy-momentum tensor of the $bc$ system is such that Virasoro central charge is $c=-26$ as in the $bc$ ghost in worldsheet string theory. 

By using the equations of motion, we can simplify them as
\begin{align}
T^{-}_{\ +} &= \frac{i}{2} \left(2 \tilde{b} \partial_+ \tilde{c} - \tilde{c} \partial_+ \tilde{b} + \tilde{\psi} \partial_+ \tilde{\psi} \right) \cr
T^{+}_{\ -} &= \frac{i}{2} \left(2 b \partial_- c  - {c} \partial_- {b} +{\psi} \partial_-\psi  \right) \cr
T^{-}_{\ -} &= \mu \tilde{b} \tilde{\psi} \left(2b\partial_- c - c\partial_-b + \psi\partial_- \psi \right) 
\cr
T^{+}_{\ +} &=  \frac{3}{2} \mu \tilde{b} \tilde{\psi} \left(2b\partial_- c - c\partial_-b + \psi\partial_- \psi \right)  \ .
\end{align}
Due to the interaction ${K}^- = i \tilde{b} \tilde{c}$ is broken
\begin{align}
\partial_- {K}^- = -2 \mu \tilde{b} \tilde{\psi}(2b\partial_- c -c\partial_- b + \psi \partial_- \psi) 
\end{align}
but the combination \eqref{twistedLorentz} is preserved.

One may improve the energy-momentum tensor so that the right-moving Virasoro is manifest
\begin{align}
\breve{T}^{+}_{\ +} = T^{+}_{\ +} - \frac{3}{4}\partial_- \tilde{K} \cr
\breve{T}^{-}_{\ +} = T^{-}_{\ +} + \frac{3}{4} \partial_+ \tilde{K} 
\end{align}
so that $\breve{T}^{+}_{+} = 0$ and $\partial_- \breve{T}^-_{\ +} = 0$. We cannot simultaneously improve the other component of the energy-momentum tensor $T^{-}_{\ -}$, but it satisfies
\begin{align}
T^{-}_{\ -} = -\frac{1}{2} \partial_- K^- 
\end{align}
so the left-moving scale symmetry does exist while there is no left-moving special conformal nor Virasoro symmetry.

As in $\mu=0$ case, we now argue that the above two theories are actually equivalent in the flat space-time. To see this, the difference between the two action (up to surface term) is 
\begin{align}
\Delta S &= S_{\mu}^{c=-26} - S_{\mu}^{c=1} \cr
&= \int dx dt \left( \mu \tilde{b}\tilde{\psi} \frac{3}{2} \partial_- (bc) \right) \ .
\end{align}
This is proportional to the equations of motion, so one should be able to remove it by field redefinition. One can see if explicitly by performing the change of variable
\begin{align}
c &\to c + 3i \tilde{\psi} bc \cr
\psi &\to \psi + \frac{3}{2} i \tilde{b} bc 
\end{align}
and see that the two actions become identical (in flat space-time).

We now sketch how to solve the equations of motions in flat space-time. Since the theory is equivalent, we work in the $c=1$ case. A systematic way to solve the equations of motion go as follows.
We begin with the right-moving part of $\tilde{b}$, $\tilde{c}$ and $\tilde{\psi}$ by specifying arbitrary functions of $x^+$
\begin{align}
\tilde{b} &= \tilde{b}(x^+) \cr
\tilde{c}_0 &= \tilde{c}_0(x^+) \cr 
\tilde{\psi}_0 &= \tilde{\psi}_0(x^+) \ .
\end{align}
They determine the conserved current $\tilde{J}$
\begin{align}
\tilde{J}(x^+) = i \tilde{b}(x^+) \tilde{\psi}_0(x^+) \ .
\end{align}
Now,we may solve 
\begin{align}
b = b(x^- + 2\mu \int^{x+} d\tilde{x}^+ \tilde{J}(\tilde{x}^+)) \cr
c = c(x^- + 2\mu \int^{x+} d\tilde{x}^+ \tilde{J}(\tilde{x}^+)) \cr
\psi = \psi(x^- + 2\mu \int^{x+} d\tilde{x}^+ \tilde{J}(\tilde{x}^+)) 
\end{align}
with arbitrary functions of $b$,$c$, and $\psi$. 

Finally, one may integrate 
\begin{align}
0&= \frac{i}{2} (2\partial_- \tilde{\psi}) - \mu \tilde{b} \left(\frac{1}{2} b \partial_- c + \frac{1}{2} c \partial_- b + \psi \partial_- \psi\right) \ ,
\end{align}
and then
\begin{align}
0 &= \frac{i}{2} \partial_- \tilde{c} + \mu \tilde{\psi} \left(\frac{1}{2} b
 \partial_- c + \frac{1}{2} c \partial_- b + \psi \partial_- \psi \right) \cr
\end{align}
to determine the inhomogeneous part of $\tilde{c}$ and $\tilde{\psi}$. Note that the homogeneous part has been already fixed by specifying $\tilde{c}_0$ and $\tilde{\psi}_0$ above. The resulting expression is consistent with our claim that the $T\bar{J}$ deformations can be regarded as the operator dependent change of coordinate
\begin{align}
x^+ &\to x^+ \cr
x^- &\to x^- - \mu \int^{x^+} d\tilde{x}^{+} J^{-} (\tilde{x}^+) \ . 
\end{align}

\section{Discussions}
In this paper, we have studied  a very special class of $T\bar{J}$ deformations of conformal field theories in two dimensions, which preserves right-moving Virasoro as well as left-moving translation and left-moving scale symmetry without left-moving special conformal symmetry. As in higher dimensional very special conformal field theories, unitarity and locality gives a no-go argument for the existence of such theories, but in this paper, we employed the twisted Lorentz symmetry to circumvent the argument. We showed concrete examples in perturbing the worldsheet string theory with the $T\bar{J}$ deformations. 

There are couple of directions to be pursued in the future. First of all, it is an interesting question if the similar construction can be performed in higher dimensions to give concrete examples of very special (conformal) field theories whose spectrum is integrable. This should give better understanding of the non-renormalizable nature of very special (conformal) field theories from the view point of the Lorentz invariant scaling. Another interesting direction is to study a holographic construction of very special $T{\bar{J}}$ deformations and its stringy embedding. At the superficial level, the model should be embedded in $SL(2) \times SL(2) \times SL(2)$ Chern-Simons theory, where two of $SL(2)$ give the duals of (original) Virasoro and the other $SL(2)$ gives the current algebra in the very special $T\bar{J}$ deformations. Finally, we have pointed out that the (very special) $T{\bar{J}}$ deformations admit two interpretations, The one is to regard the deformations as the operator dependent coordinate transformation, and the other is to regard the deformations as the operator dependent gauge transformations. The duality of the two perspectives would lead to a new way to look at the integrable nature of the $T{\bar{J}}$ deformations.

Finally, implementing $T{\bar{J}}$ deformations on the lattice is a challenging question because of the chiral nature of the interaction. It would reveal the nature of the ultraviolet completion with possibly infinite number of degrees of freedom.

\section*{Acknowledgements}
This work is in part supported by JSPS KAKENHI Grant Number 17K14301.

%%%%%%%%%%%%%%%%%%%%%%%%%%%%%%%%%%%%%%%%%%%%%%%%%%%%%%%%%%%%%%%%%%%%%%%%%%%%%%%%%%


\begin{thebibliography}{99}

%\cite{Zamolodchikov:2004ce}
\bibitem{Zamolodchikov:2004ce} 
  A.~B.~Zamolodchikov,
  %``Expectation value of composite field T anti-T in two-dimensional quantum field theory,''
  hep-th/0401146.
  %%CITATION = HEP-TH/0401146;%%
  %46 citations counted in INSPIRE as of 02 Nov 2018

%\cite{Smirnov:2016lqw}
\bibitem{Smirnov:2016lqw} 
  F.~A.~Smirnov and A.~B.~Zamolodchikov,
  %``On space of integrable quantum field theories,''
  Nucl.\ Phys.\ B {\bf 915}, 363 (2017)
  doi:10.1016/j.nuclphysb.2016.12.014
  [arXiv:1608.05499 [hep-th]].
  %%CITATION = doi:10.1016/j.nuclphysb.2016.12.014;%%
  %47 citations counted in INSPIRE as of 02 Nov 2018


%\cite{Cavaglia:2016oda}
\bibitem{Cavaglia:2016oda} 
  A.~Cavaglia, S.~Negro, I.~M.~Szecsenyi and R.~Tateo,
  %``$T \bar{T}$-deformed 2D Quantum Field Theories,''
  JHEP {\bf 1610}, 112 (2016)
  doi:10.1007/JHEP10(2016)112
  [arXiv:1608.05534 [hep-th]].
  %%CITATION = doi:10.1007/JHEP10(2016)112;%%
  %40 citations counted in INSPIRE as of 02 Nov 2018

%\cite{Dubovsky:2017cnj}
\bibitem{Dubovsky:2017cnj} 
  S.~Dubovsky, V.~Gorbenko and M.~Mirbabayi,
  %``Asymptotic fragility, near AdS$_{2}$ holography and $ T\overline{T} $,''
  JHEP {\bf 1709}, 136 (2017)
  doi:10.1007/JHEP09(2017)136
  [arXiv:1706.06604 [hep-th]].
  %%CITATION = doi:10.1007/JHEP09(2017)136;%%
  %31 citations counted in INSPIRE as of 02 Nov 2018


%\cite{Giveon:2017nie}
\bibitem{Giveon:2017nie} 
  A.~Giveon, N.~Itzhaki and D.~Kutasov,
  %``$ \mathrm{T}\overline{\mathrm{T}} $ and LST,''
  JHEP {\bf 1707}, 122 (2017)
  doi:10.1007/JHEP07(2017)122
  [arXiv:1701.05576 [hep-th]].
  %%CITATION = doi:10.1007/JHEP07(2017)122;%%
  %30 citations counted in INSPIRE as of 02 Nov 2018


%\cite{Giveon:2017myj}
\bibitem{Giveon:2017myj} 
  A.~Giveon, N.~Itzhaki and D.~Kutasov,
  %``A solvable irrelevant deformation of AdS$_{3}$/CFT$_{2}$,''
  JHEP {\bf 1712}, 155 (2017)
  doi:10.1007/JHEP12(2017)155
  [arXiv:1707.05800 [hep-th]].
  %%CITATION = doi:10.1007/JHEP12(2017)155;%%
  %19 citations counted in INSPIRE as of 02 Nov 2018

%\cite{Asrat:2017tzd}
\bibitem{Asrat:2017tzd} 
  M.~Asrat, A.~Giveon, N.~Itzhaki and D.~Kutasov,
  %``Holography Beyond AdS,''
  Nucl.\ Phys.\ B {\bf 932}, 241 (2018)
  doi:10.1016/j.nuclphysb.2018.05.005
  [arXiv:1711.02690 [hep-th]].
  %%CITATION = doi:10.1016/j.nuclphysb.2018.05.005;%%
  %19 citations counted in INSPIRE as of 02 Nov 2018

%\cite{Giribet:2017imm}
\bibitem{Giribet:2017imm} 
  G.~Giribet,
  %``$T\bar{T}$-deformations, AdS/CFT and correlation functions,''
  JHEP {\bf 1802}, 114 (2018)
  doi:10.1007/JHEP02(2018)114
  [arXiv:1711.02716 [hep-th]].
  %%CITATION = doi:10.1007/JHEP02(2018)114;%%
  %21 citations counted in INSPIRE as of 02 Nov 2018

%\cite{Kraus:2018xrn}
\bibitem{Kraus:2018xrn} 
  P.~Kraus, J.~Liu and D.~Marolf,
  %``Cutoff AdS$_{3}$ versus the $ T\overline{T} $ deformation,''
  JHEP {\bf 1807}, 027 (2018)
  doi:10.1007/JHEP07(2018)027
  [arXiv:1801.02714 [hep-th]].
  %%CITATION = doi:10.1007/JHEP07(2018)027;%%
  %22 citations counted in INSPIRE as of 02 Nov 2018

%\cite{Cottrell:2018skz}
\bibitem{Cottrell:2018skz} 
  W.~Cottrell and A.~Hashimoto,
  %``Comments on $T \bar T$ double trace deformations and boundary conditions,''
  arXiv:1801.09708 [hep-th].
  %%CITATION = ARXIV:1801.09708;%%
  %16 citations counted in INSPIRE as of 02 Nov 2018

%\cite{Baggio:2018gct}
\bibitem{Baggio:2018gct} 
  M.~Baggio and A.~Sfondrini,
  %``Strings on NS-NS Backgrounds as Integrable Deformations,''
  Phys.\ Rev.\ D {\bf 98}, no. 2, 021902 (2018)
  doi:10.1103/PhysRevD.98.021902
  [arXiv:1804.01998 [hep-th]].
  %%CITATION = doi:10.1103/PhysRevD.98.021902;%%
  %13 citations counted in INSPIRE as of 02 Nov 2018

%\cite{Babaro:2018cmq}
\bibitem{Babaro:2018cmq} 
  J.~P.~Babaro, V.~F.~Foit, G.~Giribet and M.~Leoni,
  %``$ T\overline{T} $ type deformation in the presence of a boundary,''
  JHEP {\bf 1808}, 096 (2018)
  doi:10.1007/JHEP08(2018)096
  [arXiv:1806.10713 [hep-th]].
  %%CITATION = doi:10.1007/JHEP08(2018)096;%%
  %5 citations counted in INSPIRE as of 02 Nov 2018





%\cite{Cardy:2018sdv}
\bibitem{Cardy:2018sdv} 
  J.~Cardy,
  %``The $T\overline T$ deformation of quantum field theory as random geometry,''
  JHEP10(2018)186
  doi:10.1007/JHEP10(2018)186
  [arXiv:1801.06895 [hep-th]].
  %%CITATION = doi:10.1007/JHEP10(2018)186;%%
  %22 citations counted in INSPIRE as of 02 Nov 2018

%\cite{Dubovsky:2018bmo}
\bibitem{Dubovsky:2018bmo} 
  S.~Dubovsky, V.~Gorbenko and G.~Hernandez-Chifflet,
  %``$ T\overline{T} $ partition function from topological gravity,''
  JHEP {\bf 1809}, 158 (2018)
  doi:10.1007/JHEP09(2018)158
  [arXiv:1805.07386 [hep-th]].
  %%CITATION = doi:10.1007/JHEP09(2018)158;%%
  %15 citations counted in INSPIRE as of 02 Nov 2018

%\cite{Datta:2018thy}
\bibitem{Datta:2018thy} 
  S.~Datta and Y.~Jiang,
  %``$T\bar{T}$ deformed partition functions,''
  JHEP {\bf 1808}, 106 (2018)
  doi:10.1007/JHEP08(2018)106
  [arXiv:1806.07426 [hep-th]].
  %%CITATION = doi:10.1007/JHEP08(2018)106;%%
  %11 citations counted in INSPIRE as of 02 Nov 2018

%\cite{Aharony:2018bad}
\bibitem{Aharony:2018bad} 
  O.~Aharony, S.~Datta, A.~Giveon, Y.~Jiang and D.~Kutasov,
  %``Modular invariance and uniqueness of $T\bar{T}$ deformed CFT,''
  arXiv:1808.02492 [hep-th].
  %%CITATION = ARXIV:1808.02492;%%
  %7 citations counted in INSPIRE as of 02 Nov 2018

%\cite{Bonelli:2018kik}
\bibitem{Bonelli:2018kik} 
  G.~Bonelli, N.~Doroud and M.~Zhu,
  %``$T \bar{T}$-deformations in closed form,''
  JHEP {\bf 1806}, 149 (2018)
  doi:10.1007/JHEP06(2018)149
  [arXiv:1804.10967 [hep-th]].
  %%CITATION = doi:10.1007/JHEP06(2018)149;%%
  %16 citations counted in INSPIRE as of 02 Nov 2018

%\cite{Conti:2018tca}
\bibitem{Conti:2018tca} 
  R.~Conti, S.~Negro and R.~Tateo,
  %``The $\textrm{T}\bar{\textrm{T}}$ perturbation and its geometric interpretation,''
  arXiv:1809.09593 [hep-th].
  %%CITATION = ARXIV:1809.09593;%%
  %1 citations counted in INSPIRE as of 02 Nov 2018



%\cite{Baggio:2018rpv}
\bibitem{Baggio:2018rpv} 
  M.~Baggio, A.~Sfondrini, G.~Tartaglino-Mazzucchelli and H.~Walsh,
  %``On $T\bar{T}$ deformations and supersymmetry,''
  arXiv:1811.00533 [hep-th].
  %%CITATION = ARXIV:1811.00533;%%


%\cite{SUSY}
\bibitem{SUSY}
  C.~Chang, C.~Ferko, and S.~Sethi,
  arXiv:1811.01895 [hep-th].
  %%CITATION = ARXIV:1811.01895;%%




%\cite{Guica:2017lia}
\bibitem{Guica:2017lia} 
  M.~Guica,
  %``An integrable Lorentz-breaking deformation of two-dimensional CFTs,''
  arXiv:1710.08415 [hep-th].
  %%CITATION = ARXIV:1710.08415;%%
  %26 citations counted in INSPIRE as of 02 Nov 2018


%\cite{Bzowski:2018pcy}
\bibitem{Bzowski:2018pcy} 
  A.~Bzowski and M.~Guica,
  %``The holographic interpretation of $J \bar T$-deformed CFTs,''
  arXiv:1803.09753 [hep-th].
  %%CITATION = ARXIV:1803.09753;%%
  %17 citations counted in INSPIRE as of 02 Nov 2018

%\cite{Chakraborty:2018vja}
\bibitem{Chakraborty:2018vja} 
  S.~Chakraborty, A.~Giveon and D.~Kutasov,
  %``$ J\overline{T} $ deformed CFT$_{2}$ and string theory,''
  JHEP {\bf 1810}, 057 (2018)
  doi:10.1007/JHEP10(2018)057
  [arXiv:1806.09667 [hep-th]].
  %%CITATION = doi:10.1007/JHEP10(2018)057;%%
  %10 citations counted in INSPIRE as of 02 Nov 2018


%\cite{Apolo:2018qpq}
\bibitem{Apolo:2018qpq} 
  L.~Apolo and W.~Song,
  %``Strings on warped AdS$_{3}$ via $ \mathrm{T}\bar{\mathrm{J}} $ deformations,''
  JHEP {\bf 1810}, 165 (2018)
  doi:10.1007/JHEP10(2018)165
  [arXiv:1806.10127 [hep-th]].
  %%CITATION = doi:10.1007/JHEP10(2018)165;%%
  %8 citations counted in INSPIRE as of 02 Nov 2018

%\cite{Aharony:2018ics}
\bibitem{Aharony:2018ics} 
  O.~Aharony, S.~Datta, A.~Giveon, Y.~Jiang and D.~Kutasov,
  %``Modular covariance and uniqueness of $J\bar{T}$ deformed CFTs,''
  arXiv:1808.08978 [hep-th].
  %%CITATION = ARXIV:1808.08978;%%
  %4 citations counted in INSPIRE as of 02 Nov 2018

%\cite{Cohen:2006ky}
\bibitem{Cohen:2006ky} 
  A.~G.~Cohen and S.~L.~Glashow,
  %``Very special relativity,''
  Phys.\ Rev.\ Lett.\  {\bf 97}, 021601 (2006)
  doi:10.1103/PhysRevLett.97.021601
  [hep-ph/0601236].
  %%CITATION = doi:10.1103/PhysRevLett.97.021601;%%
  %186 citations counted in INSPIRE as of 09 Jul 2017

%\cite{Cohen:2006ir}
\bibitem{Cohen:2006ir} 
  A.~G.~Cohen and S.~L.~Glashow,
  %``A Lorentz-Violating Origin of Neutrino Mass?,''
  hep-ph/0605036.
  %%CITATION = HEP-PH/0605036;%%
  %53 citations counted in INSPIRE as of 09 Jul 2017



%\cite{Nakayama:2017eof}
\bibitem{Nakayama:2017eof} 
  Y.~Nakayama,
  %``Very special conformal field theories and their holographic duals,''
  Phys.\ Rev.\ D {\bf 97}, no. 6, 065003 (2018)
  doi:10.1103/PhysRevD.97.065003
  [arXiv:1707.05423 [hep-th]].
  %%CITATION = doi:10.1103/PhysRevD.97.065003;%%
  %3 citations counted in INSPIRE as of 02 Nov 2018

%\cite{Nakayama:2018qcs}
\bibitem{Nakayama:2018qcs} 
  Y.~Nakayama,
  %``Gravity Dual for Very Special Conformal Field Theories in type IIB Supergravity,''
  Phys.\ Lett.\ B {\bf 786}, 245 (2018)
  doi:10.1016/j.physletb.2018.09.051
  [arXiv:1807.08882 [hep-th]].
  %%CITATION = doi:10.1016/j.physletb.2018.09.051;%%

%\cite{Nakayama:2018fib}
\bibitem{Nakayama:2018fib} 
  Y.~Nakayama,
  %``Local field theory construction of Very Special Conformal Symmetry,''
  Phys.\ Rev.\ D {\bf 98}, no. 2, 025007 (2018)
  doi:10.1103/PhysRevD.98.025007
  [arXiv:1802.06489 [hep-th]].
  %%CITATION = doi:10.1103/PhysRevD.98.025007;%%
  %2 citations counted in INSPIRE as of 02 Nov 2018


%\cite{Hofman:2011zj}
\bibitem{Hofman:2011zj} 
  D.~M.~Hofman and A.~Strominger,
  %``Chiral Scale and Conformal Invariance in 2D Quantum Field Theory,''
  Phys.\ Rev.\ Lett.\  {\bf 107}, 161601 (2011)
  doi:10.1103/PhysRevLett.107.161601
  [arXiv:1107.2917 [hep-th]].
  %%CITATION = doi:10.1103/PhysRevLett.107.161601;%%
  %55 citations counted in INSPIRE as of 02 Nov 2018

%\cite{Polchinski:1987dy}
\bibitem{Polchinski:1987dy} 
  J.~Polchinski,
  %``Scale and Conformal Invariance in Quantum Field Theory,''
  Nucl.\ Phys.\ B {\bf 303}, 226 (1988).
  doi:10.1016/0550-3213(88)90179-4
  %%CITATION = doi:10.1016/0550-3213(88)90179-4;%%
  %262 citations counted in INSPIRE as of 10 Jul 2017




%\cite{Sibiryakov:2014qba}
\bibitem{Sibiryakov:2014qba} 
  S.~Sibiryakov,
  %``From scale invariance to Lorentz symmetry,''
  Phys.\ Rev.\ Lett.\  {\bf 112}, no. 24, 241602 (2014)
  doi:10.1103/PhysRevLett.112.241602
  [arXiv:1403.4742 [hep-th]].
  %%CITATION = doi:10.1103/PhysRevLett.112.241602;%%
  %3 citations counted in INSPIRE as of 02 Nov 2018

\end{thebibliography}
\end{document}